\documentclass{jetpl}
\twocolumn

\lat

\title{
Measurements of anisotropic mass of magnons confined in
a harmonic trap in superfluid $^3$He-B
}

\rtitle{Measurements of magnon mass in $^3$He-B}

\sodtitle{
Measurements of anisotropic mass of magnons, confined in
a harmonic trap in superfluid $^3$He-B
}

\author{
V.~V.~Zavjalov\/\thanks{e-mail: vladislav.zavyalov@aalto.fi},
S.~Autti, V.~B.~Eltsov, P.~J.~Heikkinen}

\address{Low Temperature Laboratory,
Department of Applied Physics, Aalto University,
PO Box 15100, FI-00076 AALTO, Finland}

\rauthor{V.~V.~Zavjalov, S.~Autti, V.~B.~Eltsov, P.~J.~Heikkinen}

\sodauthor{Zavjalov, Autti, Eltsov, Heikkinen}

\abstract{We can pump magnons to a nearly harmonic magneto-textural trap in
superfluid~$^3$He-B. Using the NMR spectroscopy of levels in the trap we
have measured the anisotropic magnon mass and related values of the spin-wave
velocities. Based on our measurements we provide values of the
Fermi-liquid parameter $F^a_1$. }


\def\omegaL{\omega_\mathrm{L}}
\def\OmegaB{\Omega_\mathrm{B}}
\def\chiB{\chi_\mathrm{B}}

\def\ct{\cos\theta}
\def\st{\sin\theta}

\def\FSO{F_\mathrm{so}}

\graphicspath{{pics/}}
\newcommand{\image}[3]{
\begin{figure}[#1]
\begin{center}
\includegraphics{full_#2.eps}
\caption{\small#3}
\label{image:#2}
\end{center}
\end{figure}
}

\begin{document}

\maketitle

\section*{Introduction}

$^3$He-B is the topological superfluid with gapless Majorana fermions on
the boundary (see recent review~\cite{2015_majorana}). Optical magnons in
a magneto-textural trap proved to be a useful and convenient experimental
tool for studying various properties of superfluid $^3$He-B. A number of
effects can be observed in this system such as Bose-Einstein condensation
of magnons~\cite{BunkovVolovik2013}, Suhl instability~\cite{2009_Fomin}
with excitation of other spin-wave modes including a longitudinal Higgs
mode~\cite{OurHiggs}, self-localization of magnons~\cite{OurTrapping}. It
can be used as a probe of quantized vortices~\cite{OurVort}, Andreev
bound states and gravity waves on the $^3$He surface~\cite{OurSurf},
boundary between $^3$He-A and $^3$He-B superfluids~\cite{OurAB}. It also
can be used as a tool for accurate measurements of various $^3$He
parameters and as a thermometer which works below 0.3~mK~\cite{OurAB}.

For proper interpretation of these measurements, basic properties of
magnons should be accurately known.  In this paper we report detailed
measurements of magnon spectra from which we find the anisotropic magnon
mass and spin-wave velocity in~$^3$He-B.

\vbox{ \section*{Spin waves in $^3$He-B} }

The equilibrium state of superfluid $^3$He-B is described by the order
parameter matrix:
\begin{equation}
A_{aj}  = \frac{1}{\sqrt{3}}\ \Delta\ e^{i\varphi} R_{aj},
\end{equation}
where~$\Delta$ is the energy gap, $\varphi$ is the phase, and~$R_{aj}$
is a rotation matrix which can be written in terms of the rotation axis~$\bf n$
and the angle~$\theta$ as
\begin{equation}\label{eq:r_nt}
R^0_{a j} = \ct\ \delta_{a j} + (1-\ct)\ n_a n_j - \st\ e_{ajk} n_k.
\end{equation}
In non-zero magnetic fields the gap becomes anisotropic, but for fields
used in this work we can neglect this effect.

Spin waves in $^3$He-B correspond to oscillations of the rotation matrix
$R_{aj}$. The motion is affected by the energy of the spin-orbit
interaction $\FSO$ and the gradient energy $F_\nabla$:
\begin{eqnarray}
\label{eq:en_d}
\FSO &=& \frac{\chi_B\OmegaB^2}{15\gamma^2}  (R_{jj}R_{kk} + R_{jk}R_{kj}),\\
\label{eq:en_g}
F_\nabla
&=& \frac12 \Delta^2 ( K_1 G_1 + K_2 G_2 + K_3 G_3),
\end{eqnarray}
where
\begin{eqnarray*}
G_1 &=& \nabla_j R_{ak} \nabla_j R_{ak},\\
G_2 &=& \nabla_j R_{ak} \nabla_k R_{aj},\\
G_3 &=& \nabla_j R_{aj} \nabla_k R_{ak},
\end{eqnarray*}
$\chi_B$ is the spin susceptibility of the $^3$He-B, ${\gamma}$ the
gyromagnetic ratio for the $^3$He atom, $\OmegaB$ the Leggett frequency,
and $K_1, K_2$ and $K_3$ are parameters of the gradient energy.

The linear equation of small spin oscillations near the equilibrium value
${\bf S}^0=(\chi_B/\gamma)\ {\bf H}$ is~\cite{1983_theodorakis}
\begin{eqnarray}\label{eq:ham_eq3}
\ddot S_c &=& [\dot {\bf S}\times \gamma {\bf H}]_c + \\\nonumber
&+& \frac{\Delta^2\gamma^2}{\chi_B} \left[
K\ \nabla^2\ S_c
- K'\ \nabla_j R^0_{cj} R^0_{ak} \nabla_k\ S_a\right] - \\\nonumber
&-& \OmegaB^2\ {\bf\hat n}\cdot ({\bf S} - {\bf S^0})\ \hat n_c,
\end{eqnarray}
where~$K=2K_1+K_2+K_3$ and~$K'=K_2+K_3$.

In a texture where $\bf n$ is almost parallel to $\bf H$ or in a high
magnetic field $\omegaL = |\gamma H| \gg \OmegaB$ one can separate
transverse ($\bf S-S^0 \perp H$) and longitudinal ($\bf S-S^0 \parallel
H$) oscillations of spin. For a harmonic solution ${\bf S}-{\bf S^0} =
{\bf s}\ e^{i\omega t}$ one can write
\begin{eqnarray} \label{eq:spinwaves}
\nonumber
\left[
 - c_\perp^2\nabla^2
 - (c_\parallel^2-c_\perp^2) \nabla_j \hat l_j \hat l_k \nabla_k
+ \frac{\Omega_B^2}{2} \sin^2\beta_n
\right] s_+ = &&\\
\nonumber
= \omega(\omega-\omegaL)\ s_+,&&\\
\nonumber
\left[
 - C_\perp^2\nabla^2
 - (C_\parallel^2-C_\perp^2) \nabla_j \hat l_j \hat l_k \nabla_k
+ \Omega_B^2 \cos^2\beta_n
\right] s_z = &&\\
\nonumber
= \omega^2\ s_z, &&\\
\end{eqnarray}
where $s_+ = \frac{1}{\sqrt2}(s_x + i s_y)$, $\beta_n$ is an angle between ${\bf\hat n}$ and ${\bf H}$,
the orbital anisotropy axis $\hat l_j = R_{aj} \hat S^0_a$, and
\begin{equation}\label{eq:swvel}
\begin{aligned}
&c_\perp^2 = \frac{\gamma^2\Delta^2}{\chi_B}(K-K'/2),\quad
c_\parallel^2 = \frac{\gamma^2\Delta^2}{\chi_B} K,
\\
&C_\perp^2 = \frac{\gamma^2\Delta^2}{\chi_B} K,\quad
C_\parallel^2 = \frac{\gamma^2\Delta^2}{\chi_B} (K-K').
\end{aligned}
\end{equation}

In the case of short wavelengths (when the spin changes on a much shorter
distance than the texture) one can write spectra for plane waves with
a wave vector~$\bf k$:
\begin{eqnarray} \label{eq:spinwaves_qc}
\nonumber
 c_\perp^2\ k^2 + (c_\parallel^2-c_\perp^2) ({\bf k\cdot\hat l})^2
+ \frac12 \OmegaB^2 \sin^2\beta_n
&=& \omega(\omega-\omegaL),\\
 C_\perp^2 k^2 + (C_\parallel^2-C_\perp^2) ({\bf k\cdot\hat l})^2
+ \OmegaB^2 \cos^2\beta_n &=& \omega^2,
\end{eqnarray}

Here the meaning of all parameters becomes clear: $c_{\perp}$ and
$c_{\parallel}$ are velocities of transverse waves, propagating
perpendicular and parallel to the $\bf \hat l$ direction; $C_{\perp}$ and
$C_{\parallel}$ are the velocities of longitudinal waves; $\Omega_B$ is
a frequency of the uniform longitudinal NMR in a texture with $\bf n ||
H$.

The first equation in~(\ref{eq:spinwaves_qc}) describes transverse spin
waves, which are similar to that in ferromagnets. In the presence of
magnetic field it has two solutions $\omega(k)$, which are called
acoustic (low $\omega$) and optical (high $\omega$) magnons. The second
equation for longitudinal waves is unique for $^3$He. Spin-wave spectrum
in a uniform texture is presented on Fig.~\ref{image:quasicl}.

\image{h}{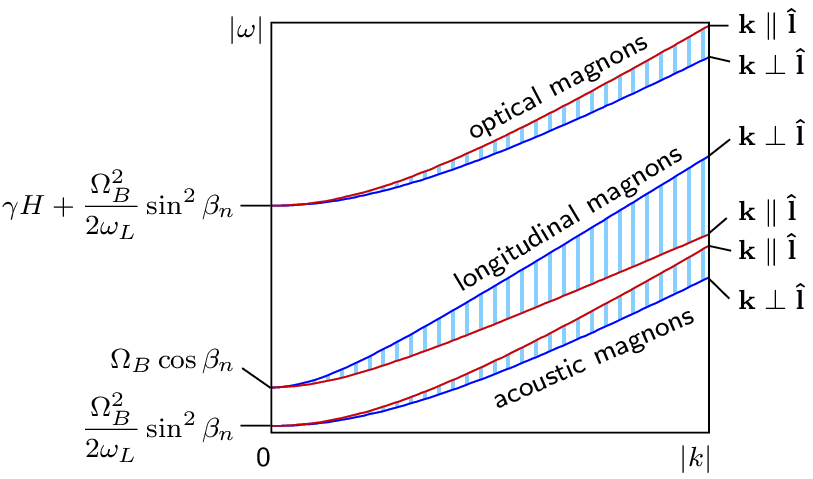}{Fig.~1. Quasiclassical spin-wave spectrum in a uniform
texture ($\beta_n=\mbox{const.}$). There are two branches of transverse waves and one of
longitudinal waves~(\ref{eq:spinwaves_qc}). Slopes of the branches at
$k\rightarrow\infty$ are spin-wave velocities, they depend on the
direction of propagation. Values at $k=0$ give resonance frequencies in the uniform NMR.}

Gradient energy coefficients were calculated in
Refs.~\cite{1975_cross,1981_dorfle}. In particular, they depend on
two antisymmetric Fermi-liquid parameters~$F^a_1$ and~$F^a_3$.
Neglecting the high-order parameter~$F^a_3$, one have:
\begin{eqnarray}\label{eq:grad_th}
K_1 = K_2 &=&
\frac{2}{\Delta^2}
\ \frac{\hbar^2 \rho}{40\ m m^*}
\ \frac{(1+\frac13 F_1^a)(1-Y_0)}{1 + \frac13 F_1^a - \frac15 F_1^a (1-Y_0)},\\
K_3 &=& K_1\ \frac{1 + \frac13 F_1^a}{1 + \frac13 F_1^a Y_0}. \nonumber
\end{eqnarray}
Here~$m$ is the bare mass of $^3$He, $m^*$ is the effective mass of
Fermi-liquid quasiparticles, $\rho$ is the $^3$He density and~$Y_0$ is
the temperature-dependent Yosida function.

Without Fermi-liquid corrections $K_1=K_2=K_3$ and one has
\begin{equation}
c_\perp/c_\parallel = \sqrt{3/4},\qquad
C_\perp/C_\parallel = \sqrt{2}.
\end{equation}

\section*{Schr\"odinger equation for optical magnons}

In the case of optical magnons with~$\omega\approx\omegaL$, in the
texture where~$\bf\hat n$ is almost parallel to~$\bf H$, the first equation
in~(\ref{eq:spinwaves}) can be rewritten in a form of a Schr\"odinger
equation for a magnon quasiparticle with an anisotropic mass:
\begin{equation}\label{eq:schred}
\left[
- \frac{\nabla_x^2+\nabla_y^2}{2m_\perp}
- \frac{\nabla_z^2}{2m_\parallel}
+ U \right] s_{+} = E\ s_{+}
\end{equation}
Here complex value $s_+$ plays role of the magnon wave function,
the energy is defined by the precession frequency~$E=\hbar\omega$,
and the values of the magnon mass are
\begin{equation}\label{eq:mmagn}
m_\perp = \frac{\hbar\omegaL}{2c_\perp^2},\quad
m_\parallel = \frac{\hbar\omegaL}{2c_\parallel^2}.
\end{equation}
Potential for magnons $U$ is formed by the order parameter
texture~$\beta_n$ and the magnetic field~$\omegaL$:
\begin{equation}\label{eq:umagn}
U = \frac{\hbar\OmegaB^2}{2\omegaL} \sin^2\beta_n + \hbar\omegaL.
\end{equation}
In our setup we are able to create a harmonic trap for magnons in
$^3$He bulk far from cell walls. Using spectroscopy
of levels in the trap we measure the magnon mass.

\section*{Experimental setup}

\image{h}{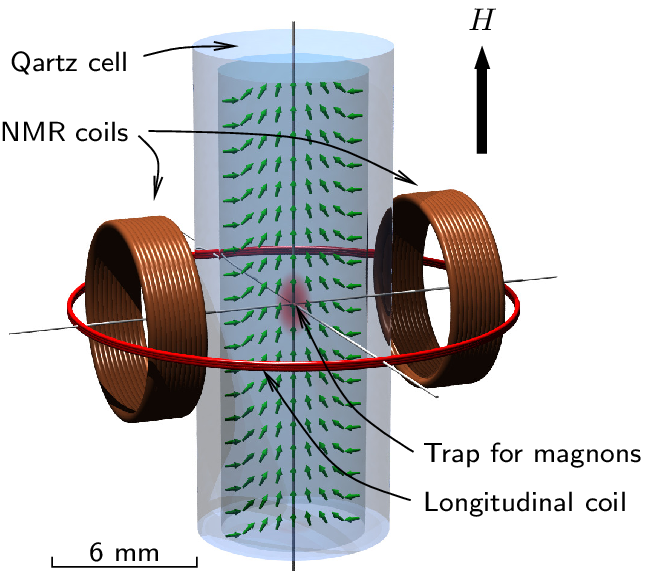}{Fig.~2. Top part of the experimental cell.
Arrows in the cell volume represent the order parameter texture ($\bf\hat l$
vector). }

We work with a $^3$He sample confined in a long quartz tube (diameter
5.85~mm, length 15~cm) and cooled in a nuclear demagnetization cryostat.
Temperature is measured by two vibrating tuning forks, located in the
lower part of the experimental cell. Experiments are performed in the low
temperature limit ($T=0.13-0.20\ T_c$), where such  parameters as the
gap~$\Delta$, Leggett frequency~$\OmegaB$, spin wave
velocities~$c_{\parallel,\perp}$, susceptibility $\chiB$ do not depend on
temperature. Pressures 0--29~bar are used.

The experimental volume is located near the upper end of the tube
(Fig.~\ref{image:cell}). The NMR spectrometer includes a transverse
pick-up coil made from copper. The coil is a part of the tuned tank
circuit. Capacitor of the circuit is installed at the mixing chamber
temperature, it can be switched to 8 different values changing the
resonance frequency in the range 550--830~kHz, which corresponds to the
NMR in $^3$He at the magnetic field 17.0--25.5~mT. The Q value of the
tank circuit is in the range 125--135 depending on the frequency.

In addition to the NMR solenoid, which produces a static magnetic field, a
small superconducting longitudinal coil is used to create a minimum of
the field at the center of the coil system. For the interpretation of
the measurements it is important to know the field profile. We determine
the profile using continues-wave (CW) NMR spectra measured in the
normal~$^3$He (Fig.~\ref{image:field_norm}).

\section*{Magnetic field profile}

\image{h}{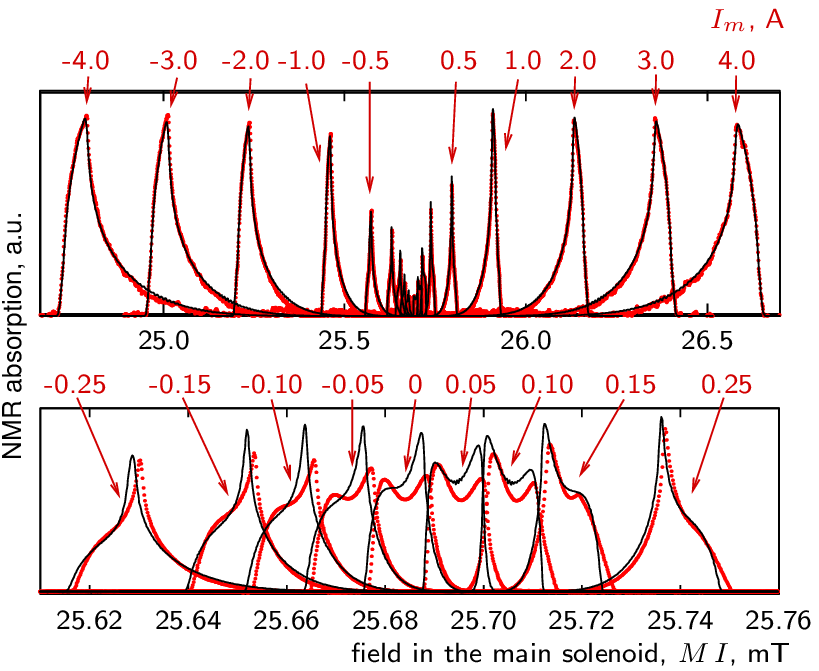}{ Fig.~3. Measured
(points) and calculated (lines) NMR signals in normal $^3$He at different
values of $I_m$ for the top spectrometer. Upper plot shows $I_m$ in the
range from $-4$~A to $4$~A, amplitude is multiplied by $|I_m|$; lower
plot shows $I_m$ in the range $-0.25$~A to $0.25$~A. }

In the simplest model the field of the main solenoid is uniform
and proportional to the current~$I$ in it. The field of the longitudinal
coil can be calculated as that of a current loop with the radius~$R_m$,
number of turns~$N_m$ and current~$I_m$.

Since both coils are superconducting, they distort the field. We have
studied this effect numerically. Distortion of the main solenoid field
can be accounted for by introduction of some additional current in the
longitudinal coil~$I^0_m \propto I$. Distortion of the longitudinal coil
field can be taken into account by introducing an additional uniform
field proportional to $I_m$ and adjusting the effective radius~$R_m$ of
the loop.

We also introduce a tiny transverse gradient $g = \partial H/\partial
x$ to explain the appearance of the double peak at~$|I_m|<0.1$~A. This
effect is small and not important for most of our measurements (since it
does not affect the quadratic terms in $H$), we use it only to improve
fitting of normal phase spectra.

The combined field is
\begin{equation}\label{eq:field_prof}
H_z = M\,I - \left[M_0 + F(r,z)-F(0,0)\right]\,I^*_m + g\ x,
\end{equation}
where $F(r,z)$ is a field of a circular loop with~$N_m$ turns, radius
$R_m$ and 1~A current and parameters have been found by fitting the
normal $^3$He CW NMR spectra:
\begin{eqnarray}\label{eq:field_profA}\nonumber
M     &=& 9.66914 \mbox{~mT/A},\quad
M^0    = 0.22305\pm0.00005 \mbox{~mT/A},\\\nonumber
I^*_m &=& I_m +  I_m^0,\quad\quad
I^0_m = - 0.0292\ I,\\
R_m   &=& 1.032 \pm 0.005 \mbox{~cm},\quad
g     = 0.02 \mbox{~mT/cm}
\end{eqnarray}
Minus sign in front of the second term in~(\ref{eq:field_prof}) shows
that the longitudinal coil is directed opposite to the NMR solenoid to
provide field minimum along the $z$ axis for a positive current~$I_m$.
Measured and calculated spectra in normal $^3$He are shown in
Fig.~\ref{image:field_norm}.

In experiments with trapped magnons only quadratic terms in the field
distribution near the center of the experimental volume are important.
Expansion of the analytical formula for a field of a current loop gives
\begin{equation}\label{eq:curr_loop}
F(r,z) = N_m\ \frac{\mu_0}{2R_m} \left(
 1 - \frac{3z^2}{2R_m^2} + \frac{3 r^2}{4R_m^2}
\right),
\end{equation}
where $\mu_0$ is vacuum permeability. Note that the
ratio of quadratic terms of~$H_z$ in the~$z$ and~$r$ direction
equals~$-2$. This comes from the Maxwell equations and should be valid for any
field distribution with this kind of symmetry. Our model for the
field profile near the center gives
\begin{equation}\label{eq:field_trap}
H_z = H_0 -  M_r (r^2 - 2z^2) I^*_m.
\end{equation}
where $M_r = 3N\mu_0/8R_m^3 = 0.1652\pm0.0024$~mT/A/cm$^2$ and $H_0$ is the
field in the center.

\section*{Magnon spectra measurement}

\image{h}{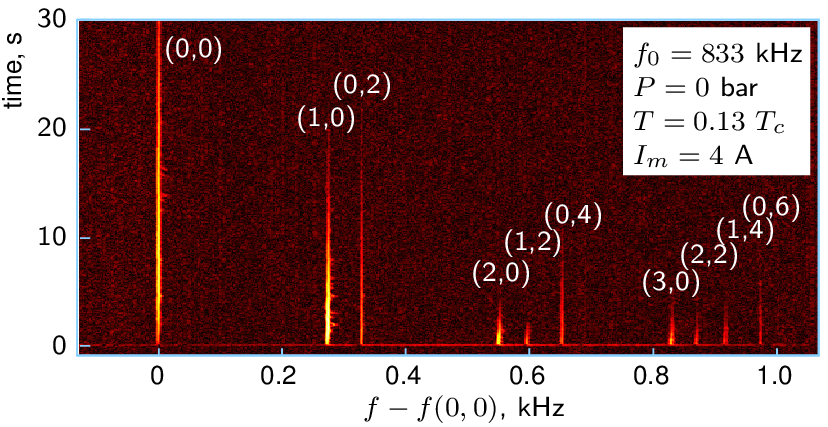}{Fig.~4. Excitation of magnon levels in pulsed NMR.
Color-coded amplitude of the Fourier transform of
the signal from the pick-up coil is plotted as a function of time.
A pulse of 0.96~ms duration excites a wide range of levels, seen as
vertical lines. Levels are marked by quantum numbers $(n_r, n_z)$.}

In our cylindrical cell order parameter of $^3$He-B forms a so-called
``flare-out'' texture~\cite{1977_flareout}.  Near the cell axis the angle
$\beta_n$ can be approximated by a linear function of radial coordinate,
$\beta_n(r) \approx \beta_n' r$. Together with the quadratic field
profile~(\ref{eq:field_trap}) this results in a harmonic trap for
magnons. The potential~(\ref{eq:umagn}) can be written in the form
\begin{equation}
U = \hbar\omega_0 + m_\parallel\frac{\omega_z^2 z^2}{2} + m_\perp\frac{\omega_r^2 r^2}{2},
\end{equation}
where
\begin{eqnarray}\label{eq:omega_rz}
&& \omega_0 = |\gamma| H_0,\qquad
\omega_z^2 = 8 c_\parallel^2 \frac{M_r I^*_m}{H_0},\qquad\\\nonumber
&& \omega_r^2 =
 2\left(\frac {c_\perp\OmegaB \beta_n'}{\gamma H_0} \right)^2
 -  4c_\perp^2 \frac{M_r I^*_m}{H_0}.
\end{eqnarray}

We can observe only axially symmetric and $z$-even eigenstates in the harmonic
potential, since they have non-zero total transverse magnetization.
Corresponding frequencies are
\begin{equation}\label{eq:resonances}
2\pi f_{(n_r,n_z)} = \omega_0 + (2n_r + 1)\ \omega_r + (n_z+1/2)\ \omega_z,
\end{equation}
with $n_r=0,1,2\ldots$ and $n_z = 0,2,4\ldots$

\image{h}{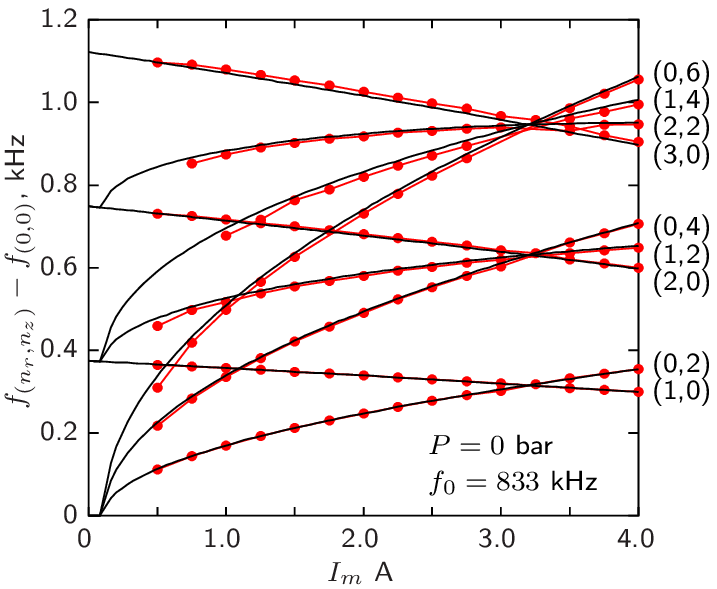}{Fig.~5. Examples of spectra
measurements. Differences between excited levels and the ground level,
$f_{(n_r, n_z)} - f_{(0,0)}$ is plotted as a function of $I_m$. Whole set
of lines is a single fit to Eq.~(\ref{eq:resonances}). Only (1,0) and
(0,2) states are used to find the fit parameters. }

We use pulsed NMR to populate a few lowest levels in this harmonic trap
(Fig.~\ref{image:exc}). If the number of magnons in the system is small
enough, interaction between the levels is negligible and multiple states
can be resolved in the measurement using Fourier transform as in
Fig.~\ref{image:exc}. We measure differences $f_{(n_r, n_z)} - f_{(0,0)}$
between higher levels and the ground level. This can be done with a
precision better than 1~Hz. The measurements are repeated as a function
of the longitudinal coil current $I_m$. With this procedure we
separate the magnetic part of the potential which depends on $I_m$ from the
textural part. In Fig.~\ref{image:spec} an example of such measurement is
presented. Data are fitted using the resonant
condition~(\ref{eq:resonances}) with frequencies~(\ref{eq:omega_rz}). We
use only $n_r=1$ and $n_z=1$ states at $I_m>0.5$~A to reduce inharmonic
effects which grow with increasing spatial extent of the standing spin wave.

\image{h}{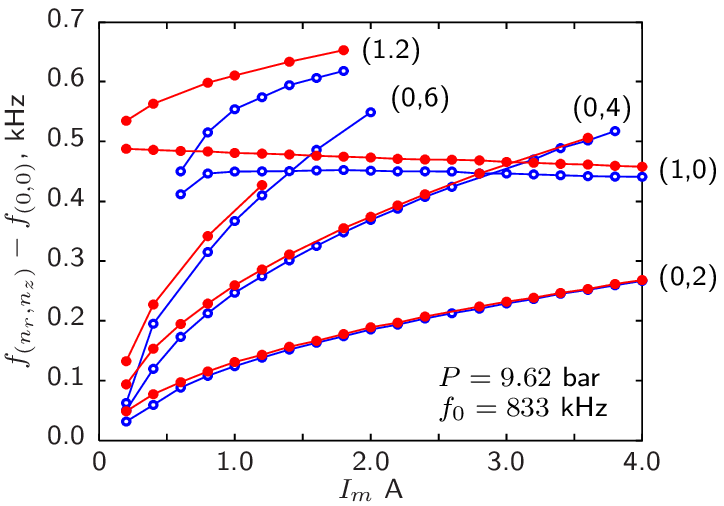}{
Fig.~6. Spectra measured in a texture with and without
defects (open and filled circles correspondingly). }

We have found that the spectrum can be affected by textural defects,
created when $^3$He is cooled down from the normal phase~\cite{1989_nmr}.
Fig.~\ref{image:spec_bad} shows the magnon spectra measured in textures with
and without defects. It is possible to remove defects by applying a large
NMR pumping for a period of a few seconds.

We have done spectra measurements for NMR frequencies $f_0 = 550-830$~kHz
and pressures $P=0-29$~bar. We fit the data using four parameters: $M_r
c^2_\parallel$, $M_r c^2_\perp$, $c_\perp\Omega_B\beta_n'$ and~$I_m^0$.
The first two parameters describe the magnetic part of the potential,
they are responsible for the axial and radial level dependence on~$I_m$.
Using these parameters and the~$M_r$ constant~(\ref{eq:field_trap}) one
can find spin wave velocities. As expected, they do not depend on the
frequency of the measurements. The pressure dependence is shown on
Fig.~\ref{image:swvel}. Accuracy of $c_\parallel$ measurement is much
better then that of~$c_\perp$ because axial levels have stronger
dependence on~$I_m$. The bad texture also has a smaller effect on it.
Theoretical values, calculated using~(\ref{eq:swvel})
and~(\ref{eq:grad_th}) are shown by dashed lines, the ratio
$c_\parallel/c_\perp$ is presented on the inset of the figure. The
smoothed measured values can be represented as (solid lines in the
figure):
\begin{eqnarray}\label{eq:swvel_appr}
c_\parallel(P\,\mbox{[bar]}) &=& \frac{387}{20.544+P} + 5.94\ \mbox{[m/s]}\\
c_\perp(P\,\mbox{[bar]}) &=& \frac{348}{21.138+P} + 5.13\ \mbox{[m/s]}
\end{eqnarray}

\image{h}{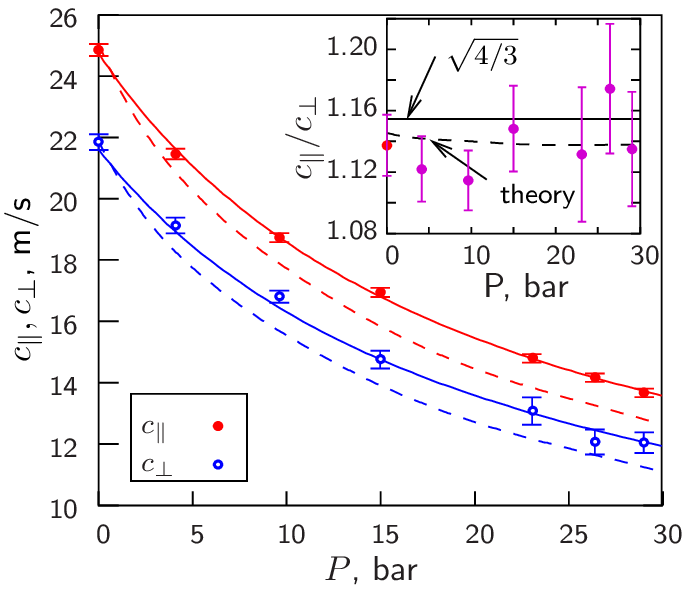}{
Fig.~7. Spin-wave velocities as a function of pressure. Filled and open
circles with error bars show measured values of~$c_\parallel$
and~$c_\perp$. Dashed lines are theoretical values~(\ref{eq:swvel}), solid
lines are approximations~(\ref{eq:swvel_appr}). In the inset the ratio
$c_\parallel/c_\perp$ is plotted.}

The third fitting parameter,~$c_\perp\Omega_B\beta_n'$ describes textural
part of the radial potential, which does not depend on~$I_m$. It gives us
information about the texture close to the cell axis. The detailed
structure of the texture obtained from the magnon spectra measurements
will be published elsewhere. The last fitting parameter,~$I_m^0$ is
proportional to the NMR frequency and does not depend on pressure. Its
value is given by~(\ref{eq:field_profA}).

\section*{Fermi-liquid parameter F\lowercase{$\bf _1^a$}}

\image{h}{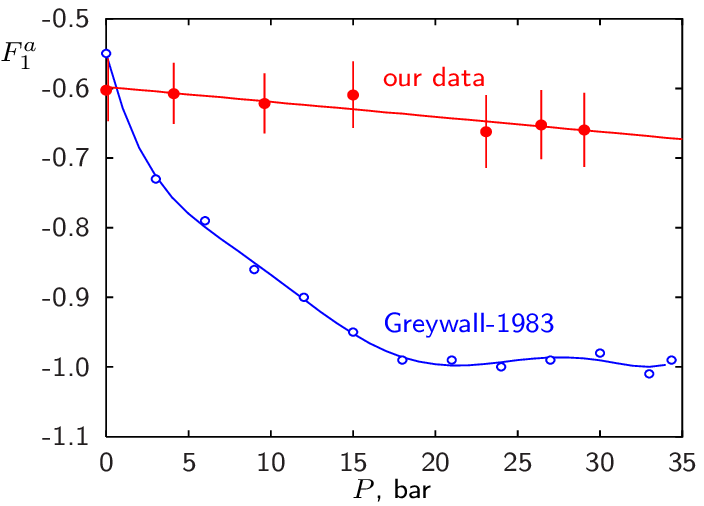}{
Fig.~8. Fermi-liquid parameter~$F^a_1$ as a function of pressure. Values
restored from our spin-wave velocity measurements
(Fig.~\ref{image:swvel}) are shown by filled circles. The error bars show
statistical uncertainty from the measurements, possible systematic error
from ignoring high-order Fermi-liquid parameters and
strong-coupling effects is not included.
 Open circles are measurements from~\cite{1983_graywall}. }

We can use expressions~(\ref{eq:swvel}) and~(\ref{eq:grad_th}) to
calculate spin-wave velocities in in the whole range of temperatures and
pressures. Values of susceptibility~$\chiB$ and effective mass~$m^*$ are
known reasonably well~\cite{VollhardtWolfle1990}. For theoretical curves
in Fig.~\ref{image:swvel} we use value of~$F^a_1$
from~\cite{1983_graywall}. It was found using normal~$^3$He specific heat
measurements at relatively high temperatures (20--100~mK), where the
Fermi-liquid approximation might not be very accurate. Higher-order
Fermi-liquid parameters were neglected in that work. Other measurements
of~$F^a_1$~\cite{1971_corruccini, 1977_osheroff} claim even less
accuracy. We have found~$F^a_1$ from our measured values of the spin-wave
velocities (see Fig.~\ref{image:f1a}). In the calculation we used
Eq.~(\ref{eq:grad_th}) which was obtained in the weak-coupling
approximation and neglects high-order Fermi-liquid parameters.
Our results can be approximated as
\begin{eqnarray}\label{eq:f1a_appr}
F^a_1 (P\,\mbox{[bar]}) = - 0.598 - 0.00214\ P.
\end{eqnarray}

This pressure dependence is much weaker than that found in
Ref.~\cite{1983_graywall}. The discrepancy at high pressures might
originate in the strong-coupling effects which are not included in the
used theoretical model.

\section*{Conclusion}

In this paper we have presented our measurements of spectra of magnons in
a magneto-textural trap in~$^3$He-B. The trap size can be controlled by
the magnetic field, which allowed us to separate the magnetic and
textural effects on the magnon levels in the trap and to measure
spin-wave velocities. Values of the spin-wave velocities determine
anisotropic magnon mass tensor and can be used to find Fermi-liquid
parameter~$F^a_1$. These new data could be used in future to refine
values of other~$^3$He-B parameters, including properties of the orbital
order-parameter texture and the magnetic relaxation properties like spin
diffusion. The latter is essential for application of trapped magnons as
a self-calibrating thermometer in a microkelvin regime.

\section*{Acknowledgements}
We thank G.E. Volovik for useful discussions. This work has been
supported in part by the Academy of Finland (project no. 250280), and by
the facilities of the Cryohall infrastructure of Aalto University. P.J.H.
acknowledges financial support from the V\"{a}is\"{a}l\"{a} Foundation of
the Finnish Academy of Science and Letters. S.A. acknowledges financial
support from the Finnish Cultural Foundation.

\end{document}